\documentstyle[amsfonts,12pt]{article}

\newcommand\func{\rm}
\renewcommand\baselinestretch{1.5} 
\begin{document}

\begin{titlepage}
\renewcommand{\baselinestretch}{1}
\renewcommand{\thepage}{}
\title{\bf A Generalized  Smoluchowsky Equation:\\
 The Hydrodynamical and Thermodynamical \\
Picture  of  Brownian Motion}
\author{L. A. Barreiro, J. R. Campanha  and R. E. Lagos \\
Departamento de F\'{\i}sica\\ 
IGCE UNESP\\
CP $178$,  $13500$-$970$ Rio Claro, SP, Brazil} 
 
\date{}
\maketitle

\begin{abstract} 
We present a systematic expansion of Kramers equation in the high friction limit. 
The latter is expanded within an operator continued fraction scheme. The relevant 
operators include  both temporal and spatial derivatives and a covariant derivative or 
gauge like operator associated to the potential energy. Trivially, the first order term yields
the Smoluchowsky equation. The second order term is readily obtained,  known as the 
corrected Smoluchowsky equation. Further terms are computed in  compact and 
straightforward fashion. As an application, the nonequilibrium thermodynamics and hydrodynamical 
schemes for the one dimensional Brownian motion is presented.

\ 

\

\begin{flushleft}
 Nonequilibrium thermodynamics, hydrodynamical equations\\

\

{\em Corresponding author}: R. E. Lagos\\
Departamento  de  F\'{\i}sica- IGCE, Universidade Estadual Paulista (UNESP)\\
C.P. 178, Rio Claro 13500-970, SP, Brazil\\
Fax: (55)-19-534-8250 \\
Email:monaco@rc.unesp.br
\end{flushleft}
\end{abstract}

\end{titlepage}

\noindent {\Large {\bf I. Introduction}}

\vspace{0.3cm}

Kramers equation \cite{kram} (although originally proposed by Klein \cite
{klein}) and the associated Smoluchowsky equation \cite{smolu} have been
widely applied to diverse problems, namely: Brownian motion in potential
wells, chemical reactions rate theory, nuclear dynamics, general stochastic
processes, just to mention a few. A relevant overview (by no means complete)
can be found in the review articles \cite{chandra}-\cite{abe}, books \cite
{schuss}-\cite{katja} and references therein. The passage from Kramers to
Smoluchowsky equation was informal and qualitative in the early days (see
for example \cite{kram} and \cite{chandra}). Only fairly recently a well
founded relationship was established \cite{dagliano}-\cite{risken3}. A
multiple time scale approach has been put forward more recently \cite{bocq1},%
\cite{bocq2}.

In section {\bf II} we present a differential recursive expansion for
Kramers equation, with respect to the inverse of the friction coefficient,
within a continued fraction formalism \cite{lagos1}- \cite{lagos3}. In a
compact and straightforward fashion we retrieve the original and the
corrected Smoluchowsky equation \cite{hanggi},\cite{gard}, in first and
third order respectively. Higher order corrections are eased out with
reasonable algebra, amenable to a diagrammatic framework. In section {\bf III%
} some applications are presented, namely the hydrodynamical and
non-equilibrium thermodynamical picture associated to the Brownian motion.
Finally, in section {\bf IV} concluding remarks include further applications
and future work.

\vspace{1cm}

\noindent {\Large {\bf II. Continued fraction expansion}}

\vspace{0.3cm}

Kramers \cite{kram} considers the probability distribution $P({\bf x,v,}t)$
in phase space for the Brownian particle (mass $m$) in an external potential 
$V({\bf x,}t)$. The friction coefficient is $\gamma $, ${\Bbb D}$ the
diffusion coefficient and $T$ the bath temperature. Kramers equation (in the
one dimensional case to simplify our analysis) is given by

\begin{equation}
\left( \frac{\partial }{\partial t}+v\frac{\partial }{\partial x}-\frac{1}{m}%
\frac{\partial V}{\partial x}\frac{\partial }{\partial v}\right) P=\gamma 
\frac{\partial }{\partial v}\left( v+\gamma {\Bbb D}\frac{\partial }{%
\partial v}\right) P  \label{kkeq}
\end{equation}

If $l$ denotes a characteristic length scale (associated to $V$ for
instance) and with Einstein relation ${\Bbb D}=\zeta k_{B}T$ \cite{einstein}
($\zeta $ is the mobility coefficient, $\zeta ^{-1}=m\gamma )$, we introduce
dimensionless variables \cite{bocq2}

\[
\tau =\frac{t}{l}\sqrt{\frac{k_{B}T}{m}},\hspace{0.5cm}q=\frac{x}{l},\hspace{%
0.5cm}\xi =v\sqrt{\frac{m}{2k_{B}T}},\hspace{0.5cm}A(q,\tau )=\frac{V(x,t)}{%
k_{B}T} 
\]

\noindent and a dimensionless friction coefficient inverse $\varepsilon $
given by

\[
\varepsilon =\frac{1}{\gamma l}\sqrt{\frac{k_{B}T}{m}}=\frac{1}{\gamma }%
\frac{v_{T}}{l} 
\]

Thus, Kramers equation is cast as

\begin{equation}
\frac{\partial }{\partial \xi }\left( \xi +\frac{1}{2}\frac{\partial }{%
\partial \xi }\right) P(q,\xi ,\tau )=\varepsilon \left( \frac{\partial }{%
\partial \tau }+\sqrt{2}\xi \frac{\partial }{\partial q}-\frac{1}{\sqrt{2}}%
\frac{\partial A}{\partial q}\frac{\partial }{\partial \xi }\right) P(q,\xi
,\tau )  \label{krameq}
\end{equation}

\noindent

Following \cite{dagliano} we use the ansatz

\begin{equation}
P(q,\xi ,\tau )=C\exp \left( -\frac{1}{2}\xi ^{2}\right) \sum_{n=-\infty
}^{\infty }\Psi _{n}(q,\tau )\Phi _{n}(\xi )  \label{ansatz}
\end{equation}

\noindent where the $\Phi _{n}(\xi )^{\prime }s$ are the orthonormal Hermite
functions \cite{sned} with $\Psi _{n}(q,\tau )\equiv \Phi _{n}(\xi )\equiv 0$
for $n<0$ and $C$ is an overall constant. Substitution yields a set of
differential recursive (difference) equations for the $\Psi _{n}\acute{}s.$
Introducing the compact notation

\[
\partial =\frac{\partial }{\partial \tau };\hspace{1cm}\nabla =\frac{%
\partial }{\partial q}\hspace{1cm}D=\nabla +(\nabla A) 
\]

\noindent we obtain the recursion relation \cite{sned}

\begin{equation}
\left( \varepsilon \partial +n\right) \Psi _{n}(q,\tau )=-\varepsilon \sqrt{n%
}D\Psi _{n-1}(q,\tau )-\varepsilon \sqrt{n+1}\nabla \Psi _{n+1}(q,\tau )
\label{recursion}
\end{equation}

This recursion relation can be readily generalized to higher dimensions via
the tensor Hermite polynomials \cite{liboff}. The first relation obtained
from equation (\ref{recursion}) with $n=0$ is a continuity equation, defined
here as the generalized Smoluchowsky equation.

\begin{equation}
\partial \Psi _{0}(q,\tau )+\nabla \Psi _{1}(q,\tau )=0  \label{smolgen}
\end{equation}

\noindent

For the $\Psi _{n}(x,\tau )$ we may define an associated function $\Psi
_{n}(x,s)$ via the transformation

\begin{equation}
\Psi _{n}(q,\tau )=\int_{\Gamma }ds\Psi _{n}(q,s)\exp (-s\tau )
\label{transf}
\end{equation}

Both the integration path $\Gamma $ and particulars of the $s$ variable are
to be defined adequately (Laplace or Fourier like), assuring among other
things that the expansion

\begin{equation}
G_{n}=-\frac{\varepsilon }{n}\left( 1+\frac{\varepsilon }{n}\partial \right)
^{-1}=\sum_{k=0}^{\infty }\left( -\frac{\varepsilon }{n}\right)
^{1+k}\partial ^{k}\hspace{1cm}n>0  \label{gexp}
\end{equation}

\noindent is a well defined expression. Here we do not require the
transformed functions, as in previous works \cite{wil}-\cite{risken3}.
Furthermore we define

\[
\alpha _{n}=\sqrt{n}DG_{n},\hspace{0.5cm}\beta _{n}=\sqrt{n+1}\nabla G_{n} 
\]

\noindent yielding a compact form for the recursion relation (\ref{recursion}%
)

\[
\Psi _{n}(q,\tau )=\alpha _{n}\Psi _{n-1}(q,\tau )+\beta _{n}\Psi
_{n+1}(q,\tau ) 
\]

With standard recursive relations and continued fractions techniques (\cite
{lagos1}-\cite{lagos3}) the above differential recursion relation can be
solved as

\begin{eqnarray}
\Psi _{n}(q,\tau ) &=&{\cal L}_{n}(\partial ,\nabla ,D)\Psi _{n-1}(q,\tau )%
\hspace{0.5cm}n>0\hspace{0.5cm}  \nonumber \\
\partial \Psi _{0}(q,\tau ) &=&-\nabla \Psi _{1}(q,\tau )=-\nabla {\cal L}%
_{1}\Psi _{0}(q,\tau )  \label{relax} \\
{\cal L}_{n}(\partial ,\nabla ,D) &=&\left( 1-\alpha _{n}{\cal L}%
_{n+1}(\partial ,\nabla ,D)\right) ^{-1}\beta _{n}  \nonumber
\end{eqnarray}

The last equation is an infinite continued fraction differential operator

\[
{\cal L}_{n}(\partial ,\nabla ,D)=\left( 1-\alpha _{n}\left( 1-\alpha
_{n+1}\left( 1-\alpha _{n+2}\left( 1-\cdot \cdot \cdot \right) ^{-1}\beta
_{n+3}\right) ^{-1}\beta _{n+2}\right) ^{-1}\beta _{n+1}\right) ^{-1}\beta
_{n} 
\]

This equation can be expanded into a sum of separable products of the kind

\begin{equation}
{\cal L}_{n}(\partial ,\nabla ,D)=\sum_{l=1}^{\infty }B_{n,l}(\nabla
,D)R_{n,l}(\partial ),\hspace{0.75cm}R_{n,l}(\partial )=\sum_{\alpha
_{1},...\alpha _{l}}C_{n,l}^{\alpha _{1}\cdot \cdot \cdot \cdot \alpha
_{l}}G_{\alpha _{1}}\cdot \cdot \cdot G_{\alpha _{l}}  \label{mix}
\end{equation}

On the other hand, from equations (\ref{smolgen}), (\ref{gexp}) and (\ref
{relax}) we obtain the identity

\begin{equation}
G_{\alpha }\Psi _{0}=-\frac{\varepsilon }{\alpha }\left( 1-\frac{\varepsilon 
}{\alpha }\nabla {\cal L}_{1}\right) ^{-1}\Psi _{0}  \label{link}
\end{equation}

Let us write the ansatz (an $\varepsilon $ expansion in purely spatial
operators)

\begin{equation}
{\cal L}_{n}(\nabla ,D)=\sum_{l=1}^{\infty }Z_{n,l}(\nabla ,D)\varepsilon
^{l}  \label{spatial}
\end{equation}

With this ansatz for $n=1$ we compute (\ref{link}), then (\ref{mix}),
obtaining an iterative relation for the $Z_{1,l}$. Once the latter are
computed we are able to compute all the ${\cal L}_{n}\acute{}s$ in terms of
purely spatial operators. We present the expansion results up to $%
\varepsilon ^{5}$ for the first five $\Psi _{n}\acute{}s$

\begin{eqnarray*}
\Psi _{1} &=&-\varepsilon \left( 1+\varepsilon ^{2}\Gamma _{3}+\varepsilon
^{4}\Gamma _{5}\right) D\Psi _{0} \\
\sqrt{2}\Psi _{2} &=&\varepsilon ^{2}\left( D+\varepsilon ^{2}D\left( \Gamma
_{3}-\frac{1}{2}\nabla D\right) +\frac{1}{2}\varepsilon ^{3}\nabla
D^{3}\right) D\Psi _{0} \\
\sqrt{3}\Psi _{3} &=&-\varepsilon ^{3}D\left( D+\varepsilon ^{2}D\left(
\Gamma _{3}-\frac{1}{2}\nabla D\right) \right) D\Psi _{0} \\
\sqrt{4}\Psi _{4} &=&\varepsilon ^{4}D^{4}\Psi _{0},\hspace{0.75cm}\sqrt{5}%
\Psi _{5}=-\varepsilon ^{5}D^{5}\Psi _{0}
\end{eqnarray*}

\noindent where

\[
\Gamma _{3}=\nabla ^{2}A,\hspace{0.75cm}\Gamma _{5}=2\left( \nabla
^{2}A\right) ^{2}+\frac{1}{2}\nabla ^{4}A+(\nabla ^{3}A)\nabla A+\frac{3}{2}%
(\nabla ^{3}A)\nabla 
\]

The generalized Smoluchowsky equation, from equation (\ref{smolgen}) and up
to $\varepsilon ^{5}$ terms is therefore

\begin{equation}
\partial \Psi _{0}=\varepsilon \nabla \left( 1+\varepsilon ^{2}\Gamma
_{3}+\varepsilon ^{4}\Gamma _{5}\right) D\Psi _{0}  \label{smolgenx}
\end{equation}

This expression coincides with previous results, see for example \cite
{risken0}. We believe to have presented a simpler and compact version for
the recursion relations, amenable to further diagrammatic manipulations.
Little attention has hitherto been paid to the evolution of moments higher
than $\Psi _{0}.$ Within an effective medium scheme \cite{lagos4} we assume $%
\Psi _{0}$ to be known via the solution of (\ref{smolgenx}) and thus compute
the higher moments $\Psi _{n}$ ($n>0$) in terms of $\Psi _{0}$. In this
fashion we explore both formal and practical consequences (the usual scheme
is to solve the recursion relations in a hierarchical form, $\Psi _{n}$ in
terms of $\Psi _{n+1}$).

\vspace{1cm}

\noindent {\Large {\bf III. Applications}}

\vspace{0.3cm}

We show that the recursive relations (\ref{recursion}) correspond to
hydrodynamic balance equations derived from the Kramers kinetic equation
(for equations other than Kramers see \cite{liboff}-\cite{kreuzer}). We may
picture the Brownian motion problem as a one dimensional fluid. Restoring to
the original (dimensional) variables, let us write some macroscopic{\bf \
densities} (moments) of interest \cite{liboff}, namely: the mass $\rho $
(particle $n$), the momentum flow $J_{m}$ (velocity density $J$ and mean
velocity $u$) , the hydrodynamic pressure $\Pi ,$ the kinetic energy $E$ and
the heat flow $J_{E}$, given respectively by (hereafter we denote $%
P(x,v,t)=P(q,\xi ,\tau ),$ given by (\ref{ansatz}))

\begin{eqnarray*}
\rho (x,t) &=&m\int dvP(x,v,t)=mn(x,t) \\
J_{m}(x,t) &=&m\int dvvP(x,v,t)=mJ(x,t) \\
&=&\rho (x,t)u(x,t),\hspace{0.75cm}u(x,t)=\frac{\int_{-\infty }^{\infty
}dvvP(x,v,t)}{\int_{-\infty }^{\infty }dvP(x,v,t)} \\
\Pi (x,t) &=&m\int dvv^{2}P(x,v,t)=2E(x,t) \\
J_{E}(x,t) &=&\frac{1}{2}m\int dvv^{3}P(x,v,t)
\end{eqnarray*}

The overall constant $C$ in equation (\ref{ansatz}) can be adjusted such
that we have the correspondence $\rho (x,t)=\Psi _{0}(x,t)$ (\ref{recursion}%
), yielding $C^{-2}=2\sqrt{\pi }mk_{B}T$. From the properties of the Hermite
functions \cite{sned} we have

\begin{eqnarray*}
\rho (x,t) &=&\Psi _{0}(x,t)=mn(x,t) \\
J_{m}(x,t) &=&v_{T}\Psi _{1}(x,t)=mn(x,t)u(x,t) \\
E(x,t) &=&\frac{1}{2}\Pi (x,t)=\frac{1}{2}v_{T}^{2}\left( \Psi _{0}(x,t)+%
\sqrt{2}\Psi _{2}(x,t)\right) \\
J_{E}(x,t) &=&\frac{1}{\sqrt{2}}v_{T}^{3}\left( \frac{3}{2}\Psi _{1}(x,t)+%
\sqrt{3}\Psi _{3}(x,t)\right)
\end{eqnarray*}

From the recursion relation (\ref{recursion}), we obtain

\noindent 
\begin{eqnarray*}
\frac{\partial \rho }{\partial t} &=&-\frac{\partial J_{m}}{\partial x}=-%
\frac{\partial \rho u}{\partial x} \\
\frac{\partial J_{m}}{\partial t} &=&-\frac{\partial \Pi }{\partial x}+K\rho
-\gamma J_{m},\hspace{1cm}K(x)=-\frac{1}{m}\frac{\partial V}{\partial x} \\
\frac{\partial E}{\partial t} &=&-\frac{\partial J_{E}}{\partial x}%
+KJ_{m}-2\gamma \left( E-E_{0}\right) ,\hspace{0.3cm}E_{0}=\frac{1}{2}%
v_{T}^{2}\rho
\end{eqnarray*}

\noindent These are balance equations for the mass, momentum and kinetic
energy density respectively. The RHS of these equations are the diffusive
(gradient), convective (proportional to the external force) and dissipative
terms, respectively. The first equation is the continuity (generalized
Smoluchowsky) equation, the only one having a conserving character (no
dissipation). There is a convective (drift) term hidden in $\Psi _{1}$. The
dissipative terms appear in a relaxation time form. The inverse relaxation
times are $0,\gamma ,2\gamma ,...n\gamma $ (as we compute higher order
terms). Here we connect to the recent multiple time scale analysis mentioned
in the introduction \cite{bocq1},\cite{bocq2}. Our formalism can be readily
generalized for multiple components undergoing chemical reactions, provided
the reaction terms are cast in the linear approximation. This is the case
for inhomogeneous semiconductors in the semiclassical approximation \cite
{mermin}-\cite{lagos5}.

We write down the Boltzmann entropy density (hereafter $k_{B}=1$)

\[
\sigma (x,t)=-\int dvP(x,v,t)\ln \frac{h}{m\func{e}}P(x,v,t)=\int dv\sigma
(x,v,t) 
\]

\noindent with $\ln \func{e}=1$. This entropy density is slightly different
to the interpolative entropy defined in \cite{kampen}. The equilibrium state
is defined as the stationary state $\partial \Psi _{n}^{\func{eq}}\equiv 0,$ 
$n=0,1,2\cdot \cdot \cdot ,$ {\bf where all flows} have ceased, {\em i.e}. $%
\Psi _{n}^{\func{eq}}\equiv 0,$ $n=1,2,\cdot \cdot \cdot $. In this context
the equilibrium state is properly defined by $D\Psi _{0}^{\func{eq}}=0$, and
we readily obtain (see \cite{kittel} for a one dimensional ideal gas)

\[
\sigma _{\func{eq}}(n_{\func{eq}}(x),T)=n_{\func{eq}}(x)\left( \frac{3}{2}%
-\ln \frac{n_{\func{eq}}(x)}{n_{Q}(T)}\right) \hspace{1cm}n_{Q}(T)=\sqrt{%
\frac{2\pi mT}{h^{2}}} 
\]

Some nonequilibrium stationary states (ss) may exist and will be reported
elsewhere. 

With the above defined entropy, our hydrodynamical picture allow us define
the local nonequilibrium {\bf densities} for the energy $E$, free energies $F
$ and $G$; the kinetic temperature $\theta $ and the internal chemical
potential $\mu _{\func{int}}$ as

\begin{eqnarray*}
E(x,t) &=&\frac{1}{2}\theta (x,t)n(x,t) \\
F(x,t) &=&E(x,t)-\theta (x,t)\sigma (x,t) \\
G(x,t) &=&F(x,t)+\Pi (x,t)=\mu _{\func{int}}(x,t)n(x,t)
\end{eqnarray*}

These definitions do not assume the local equilibrium assumption \cite
{kreuzer, vasquez} and will prove to be consistent. The total chemical
potential is $\mu =\mu _{\func{int}}+V(x)$ \cite{kittel}. From our
hydrodynamical results, and {\bf hereafter to second order in }$\varepsilon $%
, we obtain for the kinetic temperature and the particle flow density

\begin{eqnarray*}
\theta (x,t) &=&T\left( 1+\varepsilon ^{2}\frac{D^{2}n(x,t)}{n(x,t)}\right)
\\
J(x,t) &=&-\varepsilon v_{T}Dn(x,t)
\end{eqnarray*}

We estimate the temperature fluctuations $|\theta -T|/T$ for constant
external fields, for two cases at room temperature. First, for an Aluminum
Brownian particle (radius $\sim 10^{-3}$ $\func{mm}$) in water, the constant
force being gravity. The temperature fluctuation is of the order of $10^{-6}$%
. Second, for a typical semiconductor under an external electric field \cite
{mermin} $\sim 10$ $\func{V}\func{cm}^{-1}$, temperature fluctuations are of
the order $10^{-5}$. In the latter case, temperature fluctuations reach a
few percents of the bath temperature only for extremely large fields of the
order of the breakdown field $\sim 10^{5}$ $\func{V}\func{cm}^{-1}$ where
the semiclassical picture may fail for a typical semiconductor.

The entropy is given by

\begin{eqnarray*}
\sigma (x,t) &=&\sigma _{\func{eq}}(n(x,t),\theta (x,t))-\frac{\varepsilon
^{2}}{2n(x,t)}\left| \left( \frac{\partial }{\partial x}+A(x,t)\right)
n(x,t)\right| ^{2} \\
&=&\sigma _{\func{leq}}(x,t)+\sigma _{\func{gr}}(x,t)
\end{eqnarray*}

\noindent where we have separated the local equilibrium contribution from
the gradient corrections term, the latter with a Ginzburg-Landau
reminiscence \cite{landau}, here with the covariant derivative (the external
force is the affine connection). Indeed, the local equilibrium approximation
is valid up to linear corrections in $\varepsilon $. The total chemical
potential is given by

\begin{eqnarray*}
\mu (x,t) &=&\varphi (x,t)+\frac{\varepsilon ^{2}}{2}\frac{\theta (x,t)}{%
n^{2}(x,t)}\left| \left( \frac{\partial }{\partial x}+A(x,t)\right)
n(x,t)\right| ^{2} \\
\varphi (x,t) &=&\theta (x,t)\ln \frac{n^{*}(x,t)}{n_{Q}(T)}\hspace{0.7cm}%
n^{*}(x,t)=n(x,t)\exp \left( \frac{V(x)}{\theta (x,t)}\right)
\end{eqnarray*}

As expected for the total chemical potential \cite{landauer}

\[
J(x,t)=-\zeta n(x,t)\frac{\partial \mu (x,t)}{\partial x} 
\]

Finally we multiply (\ref{kkeq}) by the integrating factor $1+\ln \left( 
\frac{h}{m\func{e}}P(x,v,t)\right) $ and integrate over the velocity. We
obtain the entropy balance equation (integrating by parts, we assume $%
v^{n}P(x,v,t)\rightarrow 0,$ $v\rightarrow \pm \infty $).

\[
\frac{\partial \sigma (x,t)}{\partial t}+\frac{\partial J_{\sigma }(x,t)}{%
\partial x}=P_{\sigma }(x,t) 
\]

The computed entropy flow density $J_{\sigma }(x,t)$ (associated to entropy
exchange with the bath) and entropy production $P_{\sigma }(x,t)$
(associated to internal entropy production) are given respectively by

\begin{eqnarray*}
J_{\sigma }(x,t) &=&\int dvv\sigma (x,v,t)=\sigma (x,t)J(x,t) \\
P_{\sigma }(x,t) &=&\gamma \int dvP(x,v,t)\left( 1-\gamma {\Bbb D}\left( 
\frac{\partial \ln P(x,v,t)}{\partial v}\right) ^{2}\right) \\
&=&\frac{2\gamma }{T}\left( E(x,t)-E_{0}(x,t)\right) +2\gamma \left( \sigma
(x,t)-\sigma _{\func{leq}}(x,t)\right)
\end{eqnarray*}

The entropy production density is related to both an energy and entropy
gradient `excess' in a relaxation time fashion.

From Smoluchowsky's equation (\ref{smolgenx}) we also have the identity

\[
P_{\sigma }(x,t)=\frac{\partial n(x,t)}{\partial t}+J(x,t)\frac{\partial \ln
n(x,t)}{\partial x} 
\]

\vspace{1cm}

\noindent {\Large {\bf IV. Concluding remarks}}

\vspace{0.3cm}

The recursive method presented is compact and amenable to diagrammatic
techniques (work in progress, as an $\varepsilon $ expansion). We did show
that the moment ($\Psi $) recursion relations are nothing but the
hydrodynamical balance equations. Therefore we can readily obtain the
diffusive, convective and dissipative contributions for any relevant
quantity such as particle number, momentum and energy density.

Furthermore, we presented a nonequilibrium thermodynamical picture of
Brownian motion, hinging on the knowledge of the fundamental moment, namely
the density $\rho $, calculated via the generalized Smoluchowsky equation.

With the introduction of a generalized entropy we derived the nonequilibrium
thermodynamic density functions, the kinetic temperature and the chemical
potential. We obtain gradient corrections to the local equilibrium
approximation. We presented some relations among the computed quantities,
namely: particle flow to chemical potential and entropy gradient corrections
to entropy production.

Work in progress will probe and renormalize the thermodynamics presented
here, as higher order terms in the $\varepsilon $ expansions are considered.
Other interesting applications are related to nonisothermal oscillatory
chemical reactions \cite{lagos6, lagos7}

\vspace{0.3cm}

\noindent {\Large {\bf Acknowledgments}}

This work was supported by {\em FAPESP, SP} Brazil and {\em CNPq} Brazil.

\end{document}